\documentclass[useAMS,usenatbib,usegraphicx]{mn2e} 
\usepackage{epsfig}
\usepackage{amsmath} %need for line break in equ 3
\usepackage{rotating}           % for sideways tables/figures
\usepackage{color}     
\usepackage{graphicx}
\usepackage{times}
\usepackage{upgreek} % for up $\uptau$ 

\def\kms{km ${\rm s}^{-1}$}

\def\ch2{$\chi^2$}

\def\Lo{L$_\odot$}
\def\Mo{M$_\odot$}
\def\Moy{M$_\odot$~yr$^{-1}$}

\def\kms {\hbox{${\rm km\ s}^{-1}$}}

\def\ccm {$\hbox{{\rm cm}}^{-3}$}    %cm-3
\def\scm  {$\hbox{{\rm cm}}^{-2}$}    %cm-2
\def\arcmin {\hbox{$^{\prime}$}}
\def\MOLH {\hbox{${\rm H}_2$}}  %H2
\def \AL {$\alpha $}     %  gr. alpha
\def \HI {H{\sc \,i}}
\def \WpHz {W Hz$^{-1}$}
\def\lapp{\ifmmode\stackrel{<}{_{\sim}}\else$\stackrel{<}{_{\sim}}$\fi}
\def\gapp{\ifmmode\stackrel{>}{_{\sim}}\else$\stackrel{>}{_{\sim}}$\fi}

\title[\HI\ and OH in $z\gapp3$ CO emitters]{A search for \HI\ and OH absorption in {\boldmath  $z\gapp3$} CO emitters}\author[S. J. Curran et
al.]{S. J. Curran$^{1,2,3}$\thanks{E-mail: Stephen.Curran@vuw.ac.nz}, 
J. R. Allison$^{2,4}$,  M. T. Whiting$^{4}$,  E. M. Sadler$^{2,3}$, F. Combes$^{5}$, \newauthor M. B. Pracy$^{2}$,  C. Bignell$^{6}$
and  R. Athreya$^{7}$\\
$^{1}$School of Chemical and Physical Sciences, Victoria University of Wellington, PO Box 600, Wellington 6140, New Zealand\\
$^{2}$Sydney Institute for Astronomy, School of Physics, University of Sydney, NSW 2006, Australia\\
$^{3}$ARC Centre of Excellence for All-sky Astrophysics (CAASTRO)\\
$^{4}$CSIRO Astronomy and Space Science, PO Box 76, Epping NSW 1710, Australia\\
$^{5}$l'Observatoire de Paris, 61 Av. de l'Observatoire F-75 014 Paris, France\\
$^{6}$National Radio Astronomy Observatory, P.O. Box 2, Rt. 28/92 Green Bank, WV 24944-0002, USA\\
$^{7}$Indian Institute of Science Education and Research, 900, NCL Innovation Park, Dr Homi Bhabha Road
Pune, Maharashtra 411008, India}
 
\begin{document}

\date{Accepted ---. Received ---; in original form ---}

\pagerange{\pageref{firstpage}--\pageref{lastpage}} \pubyear{2016}

\maketitle

\label{firstpage}
\begin{abstract}
  We present the results of a survey for \HI\ 21-cm and OH 18-cm absorption in seven strong CO emitters at $z\gapp3$.
  Despite reaching limits comparable to those required to detect 21-cm absorption at lower redshifts, we do not detect
  either transition in any of the objects searched. We believe that this is due to the high redshift selection causing
  all of our targets to have ultra-violet luminosities above the critical value, where all of the atomic gas in the host
  galaxy disk is suspected to be ionised. However, not only are all of our targets bright in CO emission, but detection
  of CO above the critical UV luminosity is generally not uncommon. This suggests that the molecular gas is shielded
  from the radiation or is physically remote from the source of the continuum emission, as it appears to be from CO
  observations of high redshift radio galaxies.
\end{abstract}

\section{Introduction}
\label{intro}

Absorption of the background radio continuum by neutral hydrogen (\HI) traces the cool component of the neutral 
gas in galaxies, the reservoir for star formation. High redshift observations of the 21-cm
transition can address several pressing problems in cosmology and fundamental physics, such as the evolution of
large-scale structure (e.g. \citealt{rab+04}), detecting the Epoch of Re-ionisation (e.g. \citealt{cgfo04}) and measuring
the contribution of the neutral gas content to the mass density of the Universe (\citealt{kpec09,cur09a}). Furthermore,
spectral lines at high redshift can provide a measure of the values of the fundamental constants of nature at large
look-back times (see \citealt{cdk04}). Of these, the hydroxyl radical (OH) is of particular interest since several
different combinations of the constants can be measured from this single species \citep{dar03}. That is, OH
can potentially yield several measurements from a single sight-line, thus eliminating any relative velocity offsets
 which  would cause a change in the redshifts, thus  imitating a variation in the constants.

%FLASH/MgII/new           sort -n -k2 DLA+MgII-det.dat
%FLASH/associated        all_photo-det.dat
However, the detection of high redshift \HI\ absorption is currently very elusive, with only three detections at
$z\gapp3$ (look-back times of $\gapp11.5$ Gyr), two of which occur in galaxies intervening the sight-lines to more
distant quasars \citep{kcl06,sgp+12} and one within the host galaxy of the quasar itself \citep{ubc91}.  The former
``intervening'' systems arise in damped Lyman-$\alpha$ absorption systems (DLAs)\footnote{Defined by their large neutral
  hydrogen column densities (in excess of $N_\text {\HI}=2\times10^{20}$ \scm), detected through the Lyman-\AL\
  transition (usually redshifted into the optical band at $z\gapp1.7)$.}, with the low detection rate at high redshift
being due to the geometry of an expanding Universe, where the high redshift cases cannot occult the background emission
as effectively as those at lower redshift \citep{cur12}. In the case of molecular absorption, the optical selection
of targets means that the background quasi-stellar object (QSO) is not sufficiently obscured by the DLA \citep{cwc+11}
to indicate a suitable environment for detecting molecular gas in absorption, either by OH \citep{cwm+06} or by
molecules with transitions in the millimetre band (e.g. CO, HCO$^+$, HCN, \citealt{cwc+11}).  Thus, given that an optical
redshift biases against a dusty, obscured object, in the millimetre and decimetre bands, the number of redshifted
molecular absorption systems remains a paltry five, all of which are at a redshifts of $z\leq0.89$
(\citealt{wc94,wc95,wc96b,wc98,cdn99,kc02a,kcdn03,kcl+05}).\footnote{Note, however, that the Lyman and Werner
  ultra-violet bands of \MOLH\ have been detected in 24 DLAs (compiled in \citealt{sgp+10} with the addition of
  \citealt{rbql03,fln+11,gnp+12,sgp+12,nsr+15}) although these have molecular abundances which are generally much lower
  than those detectable in the microwave band (\citealt{cmpw03,cwc+11}).}
% papers/65/ awk < H2-list.dat '{if ($18 ~/a/) print $18}' | wc

For the ``associated'' systems, the use of an optical redshift to which to tune the receiver also introduces a bias
against the detection of neutral atomic and molecular gas, particularly at high redshift. From a survey for \HI\ 21-cm
and millimetre-band (CO, HCO$^+$ \& HCN) absorption at $z\gapp3$, \citet{cww+08} obtained zero detections in the ten
objects searched. Upon a detailed analysis of the photometry, they found all of the radio sources to be above a $\lambda
= 912$~\AA\ luminosity of $L_{\rm UV} \sim 10^{23}$ \WpHz, where 21-cm has {\em never} been detected, no matter
the selection criteria or redshift \citep{cww+08,cwm+10,cwsb12,cwt+12,gd11,ace+12,gmmo14,akk16}. This critical value
applies at all redshifts \citep{cw10} and, while 21-cm absorption is readily detected at $z\lapp1$
(e.g. \citealt{vpt+03}), above these redshifts the optical selection of targets biases towards objects which are most UV
luminous in the quasar rest-frame. In these, the \HI\ ionising ($\lambda \leq 912$~\AA) photon rate of
$Q_\text{\HI}\gapp3\times10^{56}$~sec$^{-1}$ is sufficient to ionise {\em all} of the neutral atomic gas in a large
spiral galaxy \citep{cw12}.

This suggests that even the Square Kilometre Array will not detect atomic or molecular absorption in the $z\gapp3$ radio
galaxies and quasars currently known (see figure 4 of \citealt{msc+15}). Thus, in order to find such absorption at high
redshift, it is necessary to dispense with the traditional reliance upon optical redshifts and perform spectral scans
towards optically faint radio sources.  Currently, however, sufficiently wide spectral scans are subject to severe radio
frequency interference (RFI).\footnote{For example, 200 \& 800 MHz wide scans for \HI\ and OH with the Green Bank
  Telescope (see \citealt{tcw+12}).} Therefore, in order to avoid the optical selection of targets, while retaining
relatively narrow bandwidths, we can tune the receiver to the redshift of CO emission, which has been detected over 150
times at $z\gapp0.1$. Here we present the results of the first survey for \HI\ and OH absorption in a sample of
$z\gapp3$ CO emitters.

\section{Observations and analysis}
\label{observations}

\subsection{Sample selection}

\begin{figure*}
\centering \includegraphics[angle=-90,scale=0.65]{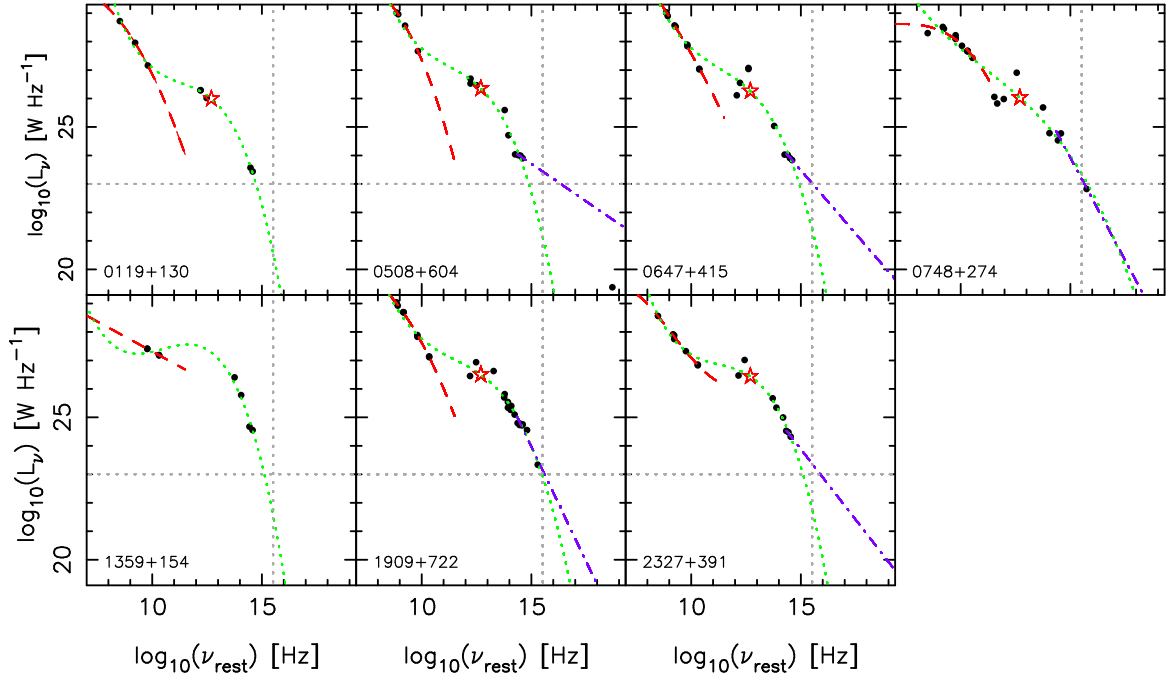}
\caption{The rest-frame SEDs of our targets overlaid by fits to the photometry. The broken curve shows the third order polynomial fit to the radio data,
the dotted curve the fit to all of the data (with the unfilled star showing the far infrared luminosity, $L_{\rm FIR}\approx5 - 14\times10^{12}$\,\Lo, estimated from this)
and the  dash--dot line the power-law fit to the UV data,
where possible (see \citealt{cwsb12} for details). The vertical dotted line shows a rest-frame frequency of $3.29\times10^{15}$ Hz ($\lambda = 912$~\AA) and the horizontal line 
the ``critical'' UV luminosity of  $L_{\rm UV}=10^{23}$ \WpHz.}
\label{SEDs_full}
\end{figure*}

From the known $z\gapp0.1$ CO emitters\footnote{\label{co_ref}Compiled from
  \citet{opg+96,dnw+99,fis+99,prv+00,cod+02,dnm+03,gip03,ngi+03,whdw03,btg+04,hsy+04,ddn+05,gbs+05,kes+05,kns+05,sv05,itn+06,tnc+06,mnb+07,csn+07,wmr07,awd+08,ccn+09,dds+09,wid+09,bct+10,dbw+10,lcs+10,rcc+10,cgb+11,dss+11,tgn+10,ytf+10,fhb+11,rie11a,efm+11,lsa+11,mds+12,msl+12,bbb+13,cgb+13,bsc+13,slg+13,tng+13,enf+14,gtl+15}.}  we selected those for which the 21-cm transition is redshifted
into the 90-cm bands of the Green Bank Telescope (GBT) and Giant Metrewave Radio Telescope (GMRT). As described in
\citet{cwsb12}, we compiled the photometry\footnote{From the NASA/IPAC Extragalactic Database (NED) and matching the
  sources from the Wide-Field Infrared Survey Explorer (WISE), Two Micron All Sky Survey (2MASS) and GALEX databases.}
and selected those with an appreciable radio flux (estimated to exceed $\approx0.2$ Jy at the redshifted 21-cm
frequency), prioritised by lowest $\lambda = 912$~\AA\ luminosity (Fig. \ref{SEDs_full}).
This yielded a shortlist of seven sources, of which six could be observed with the GBT and five with the GMRT. In order
to offer some redundancy against RFI, we targeted the four sources which could be observed by both telescopes.  Although
we attempted to select the most optically faint sources, the high redshift ($z\gapp3$) selection did nevertheless lead
to the seven shortlisted sources having $\lambda = 912$~\AA\ luminosities above (but close to) the critical value of
$L_{\rm UV}\sim10^{23}$ \WpHz\ (see Table~\ref{obs}).
 
\subsection{GMRT observations and analysis}

 Each of the sources was searched for \HI\ 21-cm absorption on 30--31 August 2013 with the GMRT full 30 antenna array, using the 325 MHz receiver
backed with the FX correlator over a bandwidth of 4 MHz.  This was spread over 512 channels in orthogonal circular
polarisations ({\sf LL} \& {\sf RR}), giving a channel spacing of $\approx7$ \kms. 
 For bandpass calibration 3C\,48, 3C\,147 and 3C\,298 were used, with the phases being self calibrated apart from
NVSS\,J012142+132058, which used  3C\,49 and B2\,2327+39.
The data were flagged and reduced using the {\sc miriad} interferometry reduction package, with
removal of the non-functioning antenna 18, leaving 406 baseline pairs.
After averaging the two polarisations, a spectrum was extracted from the cube.
Regarding each source:\\
{\bf 0119+130 (NVSS\,J012142+132058)} was observed for a total of 1.38 hours at a central frequency of 314.25 MHz. Flagging of the worst RFI left 342
baseline pairs.  The source was unresolved by the $14.6"\times10.3"$ synthesised beam.\\
{\bf B2\,0748+27} was observed for a total of 0.66 hours at a central frequency of 337.55 MHz. Only one baseline pair required flagging due to 
excessive RFI, although some ripple on the bandpass was still evident (Fig. \ref{spectra}). The source was unresolved by the $16.7"\times9.9"$ synthesised beam.\\
{\bf 1359+154 (87GB\,135911.5+152747)} was observed for a total of 2.25 hours at a central frequency of 335.01 MHz. RFI was relatively severe, and flagging all baseline
pairs above an r.m.s. noise level of 1~Jy, left just 127 pairs. 
The source was unresolved by the $29.1"\times16.0"$ synthesised beam.\\
{\bf 1909+722  (4C\,+72.26)} was observed for a total of 0.83  hours at a central frequency of 313.42 MHz, with only the flagging of one bad channel being required.
The source was unresolved by the $19.7"\times12.3"$ synthesised beam.\\
{\bf B2\,2327+39}  was observed for a total of 1.47 hours at a central frequency of 346.95 MHz. The presence of a strong bandpass ripple required flagging of the
worst affected baseline pairs, leaving 357. The source was unresolved by the $13.7"\times9.3"$ synthesised beam.\\
 
\subsection{GBT observations and analysis}

Each of the sources targeted with the GBT were observed for a total of two hours with the observations being completed
over several sessions in 2013 (ending in October). For all observations, the Prime Focus 1 (PF1) receiver was used,
backed by the GBT spectrometer, with a 12.5MHz band over 4\,098 or 8\,196 lags. This gave a channel spacing of 3.052 or 1.526
kHz, corresponding to a spectral resolution of $\approx1 - 3$ \kms, which was redressed to 10 \kms\ after averaging the
good scans.\footnote{The line-widths of the current $z\gapp0.1$ associated 21-cm detections range from 18 to 475 \kms, with a mean
  of 167 \kms\ \citep{cwsb12}.}  Four separate IFs were employed in two orthogonal linear polarisations ({\sf XX} \&
{\sf YY}), allowing us to observe both the \HI\ 21-cm and OH 18-cm (main -- 1665 \& 1667 MHz, satellite -- 1612 \& 1720 MHz)
lines simultaneously in cases where these were redshifted into the band (290--395 MHz).

Many of the observational scans were marred by RFI and those completely dominated by interference were removed. Nevertheless,
there remained RFI spikes peppered throughout the band. Given that the GBT is a single dish telescope, 
thus not giving the option of removing badly affected baseline pairs, 
there was little we could do to improve the data, although we did apply the following steps:
\begin{enumerate}
\item We fitted and removed a low order polynomial to the bandpass, although some ripple was still apparent in some
  spectra (e.g. 0748+274, Fig. \ref{spectra}).  We did not use high order polynomials in order to avoid over-fitting the
  data and thus possibly removing any putative broad, weak absorption features.
    \item Given that the detection of 21-cm emission at $z\gapp0.2$ is beyond the capabilities of current radio telescopes (e.g. \citealt{cc15}), any positive flux spikes were deemed to be caused by RFI. The noise levels and redshift ranges are quoted between these spikes (Table \ref{obs}), where they
              bracket the expected absorption frequency (shown by the downwards arrows in Fig. \ref{spectra}).
      \item Since we are searching for absorption, negative flux spikes cannot automatically be attributed to RFI, although most were similar
        in appearance to the positive flux spikes.  We distinguished the RFI from any putative absorption features by:
        \begin{enumerate}
        \item Checking whether the features were dominant in some individual scans while being absent or frequency shifted in others.  Often, as for the positive flux spikes, the spikes would be intermittent or
          dominated by a single scan (e.g. the  299 MHz feature in the \HI\ bands of 0508+604 and 0647+415).
         \item Checking if the feature resembles what we would expect. For example, for the ``absorption'' at 347.5 MHz towards 0647+415, we would expect a 
           second feature separated by $\approx0.4$ MHz, due to the OH main-line doublet. Also, \HI\ absorption  is not apparent at the expected 296.0 MHz.
           \item Checking if the feature is of the expected strength. For example, if the 381 MHz feature towards 1359+154 were real, we would expect
             the stronger main-line doublet to be apparent at $\approx394$ MHz, as well as \HI\ absorption at  335.7 MHz, none of which are  seen
             at these relatively clean frequencies.
\end{enumerate}
\end{enumerate}
Regarding the observation and data reduction for each individual target:\\
{\bf 0508+604 (4C\,+60.07)} was observed in four IFs for \HI\ (centered on 296.66 MHz), the OH 1612 MHz line (336.72
MHz), the 1665 and 1667 MHz main lines in a single IF (redshifted to 347.83 and 348.24 MHz, respectively) and 1720 MHz
(centred on 359.34 MHz).  The \HI\ band was very clean with 1.95 hours of integration on source remaining after
flagging. In the final averaged spectrum there is a prominent absorption feature apparent at 298.73 MHz (blueshifted
from the CO emission by 2060 \kms). However, the fact that this feature was not prominent in all scans, as well as being
apparent as emission in two (out of 53 scans, as well as in the 0647+415 spectrum), leads us to conclude that this is instrumental in nature. The data were
also relatively RFI-free in the other bands, leaving 1.91 hours on
source for 1612 MHz, 1.83 hours for the OH main line band and 1.99 hours for the 1720 MHz band.\\
{\bf 0647+415 (4C\,+41.17)} was observed in four IFs for \HI\ (centered on 296.16 MHz), the OH 1612 MHz line (336.16
MHz), the 1665 and 1667 MHz main lines in a single IF (redshifted to 347.25 and 347.66 MHz, respectively) and 1720 MHz
(centred on 358.74 MHz).  Flagging in the \HI\ band left a total of 0.40 hours of integration on source, 0.52 hours for
the OH 1612 MHz band, 0.55 hours for the OH main line band and 0.24 hours in the 1720 MHz band, in which all of the {\sf
  XX} polarisation was lost.  Note that the measured fluxes varied considerably between the two polarisations,
particularly for the 1612 MHz band, where for {\sf XX} this was $\approx3$ Jy, cf. $\approx1$ Jy in the {\sf YY}, with
the latter being the more consistent with the other bands and a flux density of $\approx1.6$ Jy being expected
from a fit to the NED radio photometry.\\
{\bf B2\,0748+27} was observed in two IFs for \HI\ (centered on 337.55 MHz) and OH 1612 MHz (383.13 MHz).
Some RFI was present, particularly in the lower frequency IF, and some flagging of the data was required. This left a total on source time of 0.68 hours
for \HI\ and 1.23 hours for OH, although some spikes still remained in the spectra (Fig. \ref{spectra}).\\
 {\bf 87GB\,135911.5+152747} was observed in three IFs for \HI\ (centered on 335.01 MHz), the OH 1612 MHz line (380.24 MHz) and 
the 1665 and 1667 MHz main lines in a single IF (redshifted to 392.78 and 393.25 MHz, respectively). 
After flagging of the worst RFI affected scans, the total on source integration times were 1.82 hours (\HI), 0.92 hours (OH 1612 MHz) and 1.55 hours (OH main lines).
For each scan a negative flux was noted in both polarisations. These are believed to be caused by confusion within the beam/sidelobes/off-position,
due to the large beam ($\sim40\arcmin$ at 335 MHz, \citealt{rie12}). \\
{\bf 1909+722 (4C\,+72.26)} was observed in four  IFs for \HI\ (centered on 313.42  MHz), the OH 1612 MHz line (355.74 MHz), 
the 1665 and 1667 MHz main lines in a single IF (redshifted to 367.48  and 367.91 MHz, respectively) and 1720 MHz (centred on 379.64  MHz).
Flagging left a total integration time of 1.51 hours in the \HI\ band, 1.47 hours in the  OH 1612 MHz band and 1.3 hours in the 1720 MHz OH band.
RFI affected  the middle of the main line OH band in the latter of two separate observing runs and flagging of this left 0.48 hours of integration time.\\ 
{\bf B2\,2327+39}  was observed in two IFs for \HI\ (centered on 346.95 MHz) and OH 1612 MHz (393.80 MHz). After flagging of the worst RFI
affected scans, leaving 1.19 hours of total integration time at the lower frequency, spikes were still present throughout the bandpass, which 
was clean between these (with an r.m.s. noise level of $\approx4$ mJy). Unfortunately, one of these lies at the frequency where the \HI\ absorption
is expected and so we do not quote an optical depth limit (Table \ref{obs}). Spikes were all present at the higher frequency and, although close to
the expected OH 1612 MHz absorption, 393.80 MHz is relatively clean, allowing us to quote a limit. Note that, the flux density at both frequencies
varied between the two polarisations and two separate observing sessions and so the mean values quoted should not be deemed as reliable
as that obtained from the interferometric  (GMRT) observations.

\section{Results and discussion}
\subsection {Observational results}

\begin{figure*}
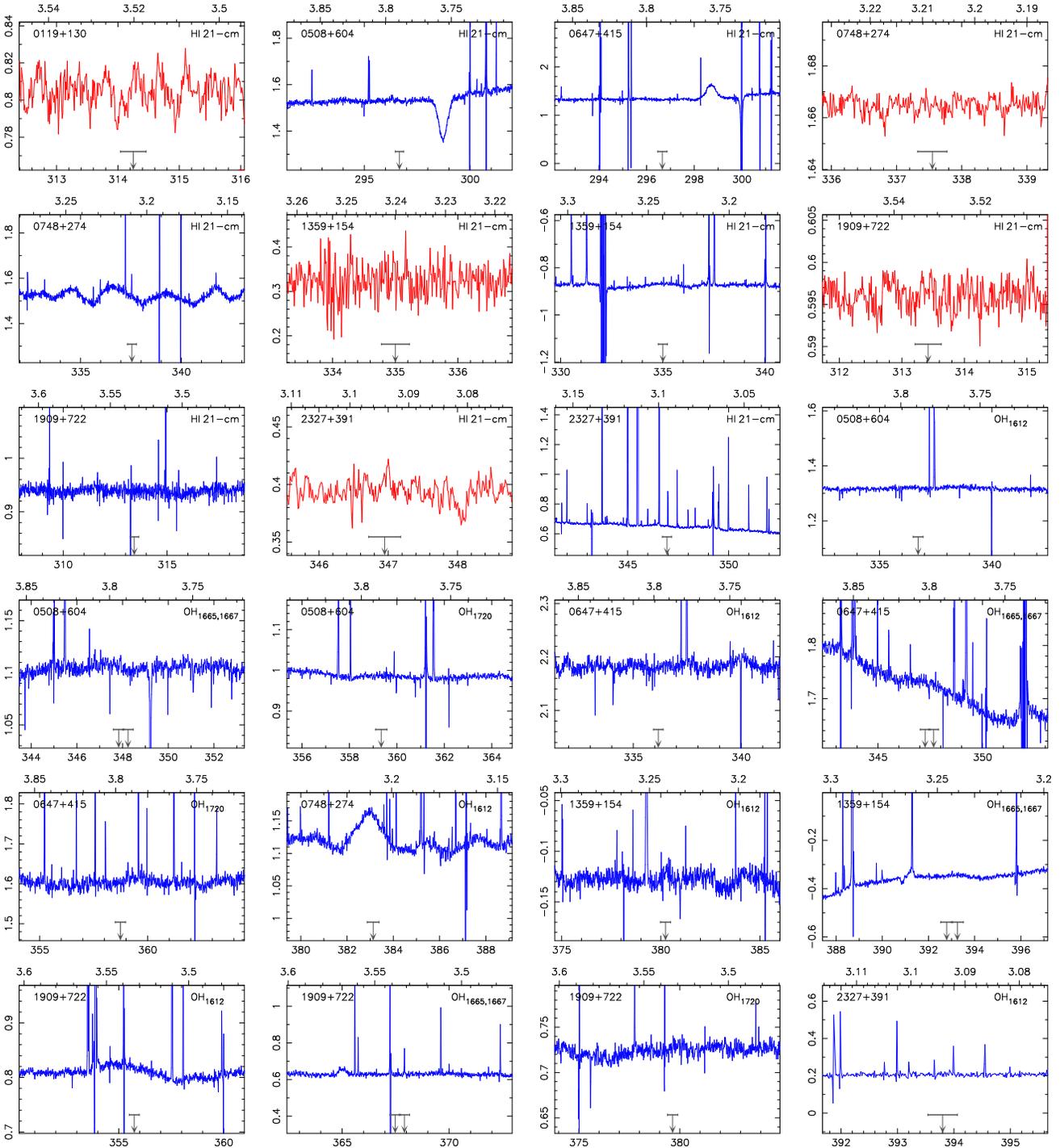

\vspace{18.7cm}   % HI spectra first
\includegraphics{tnj0121+13.314_HI_1Jy-1.dat-freq_poly9_flux_10kms.ps}
\includegraphics{4C60.07_IF0_297MHz-10kms.dat-freq_poly1_flux_9kms.ps} % 4C\,+60.07 GBT
\includegraphics{4C41.17_IF0_296MHz-10kms.dat-freq_poly0_flux_10kms.ps} % 4C\,+41.17 GBT  
\includegraphics{0748+27_HI_1.dat-freq_poly9_flux_10kms.ps} % 0748+274  GMRT
% new line
\includegraphics{0748+27_IF0_338MHz-10kms.dat-freq_poly9_flux_10kms.ps} % 0748+274 GBT
\includegraphics{1359+154_HI-to1Jy+blflag_2.dat-freq_poly6_flux_10kms.ps}  %1359+154 GMRT
\includegraphics{1359+1527_IF0_335MHz-10kms.dat-freq_poly0_flux_11kms.ps}%1359+154 GBT
\includegraphics{4c+72.26_HI_2.dat-freq_poly3_flux_10kms.ps} %4C+72.26 GMRT
%new line
\includegraphics{4C72.26_IF0_313MHz-10kms.dat-freq_poly3_flux_10kms.ps}  %4C+72.26 GBT
\includegraphics{2327+39_HI_2_to0.5Jy.dat-freq_poly3_flux_10kms.ps} %2327+39 WITH GBT
\includegraphics{2327+39_IF0_347MHz-10kms.dat-freq_poly0_flux_11kms.ps}
\includegraphics{4C60.07_IF2_337MHz-10kms.dat-freq_poly1_flux_11kms.ps} % 4C\,+60.07 1612
%new line
\includegraphics{4C60.07_IF1_348MHz-10kms.dat-freq_poly1_flux_11kms.ps} % 4C\,+60.07  main      - ref freq 348.24
\includegraphics{4C60.07_IF3_359MHz10kms.dat-freq_poly2_flux_10kms.ps} % 4C\,+60.07 1720
\includegraphics{4C41.17_IF2_336MHz-10kms.dat-freq_poly9_flux_11kms.ps} % 4C\,+41.17     1612
\includegraphics{4C41.17_IF1_347MHz-10kms.dat-freq_poly0_flux_11kms.ps}% 4C\,+41.17  main  - ref freq 347.66
%new line
\includegraphics{4C41.17_IF3_359MHz-10kms.dat-freq_poly2_flux_10kms.ps} % 4C\,+41.17  1720
\includegraphics{0748+27_IF1_383MHz-10kms.dat-freq_poly9_flux_10kms1.ps} % 0748+274  1612
\includegraphics{1359+1527_IF3_380MHz-10kms.dat-freq_poly3_flux_10kms.ps} % 1359+154  1612
\includegraphics{1359+1527_IF1_393MHz-10kms.dat-freq_poly2_flux_9kms.ps}% 1359+154  main     - ref freq  393.25
%new line
\includegraphics{4C72.26_IF2_356MHz-10km.dat-freq_poly3_flux_10kms.ps}% 4C\,+72.26  1612
\includegraphics{4C72.26_IF1_367MHz-10kms.dat-freq_poly2_flux_10kms.ps}% 4C\,+72.26  main  - ref freq   367.91 
\includegraphics{4C72.26_IF3_380MHz10kms.dat-freq_poly4_flux_10kms.ps} % 4C\,+72.26 1720
\includegraphics{2327+39_IF1_394MHz-10kms.dat-freq_poly0_flux_10kms.ps}    %        2327+391  1612
 \caption{The final averaged \HI,  followed by the OH spectra (the 1665 \& 1667 MHz main line spectra are shown in
  a single plot). The GBT spectra are shown in blue with the GMRT in red. The ordinate shows the  flux density [Jy] and the abscissa the
  barycentric frequency [MHz] at a resolution of 10 \kms. The scale
  along the top axis shows the redshift for the appropriate transition (1667 MHz in the case of the OH main line
  spectra). The downwards arrow shows the expected frequency of the
  absorption from the CO redshift, with the horizontal bar showing a span of $\pm200$ \kms\ for guidance.}
\label{spectra}
\end{figure*}

\begin{figure*}
\vspace{29.0 cm} \setlength{\unitlength}{1in} 
\begin{picture}(0,0)
\put(-4,1){\includegraphics{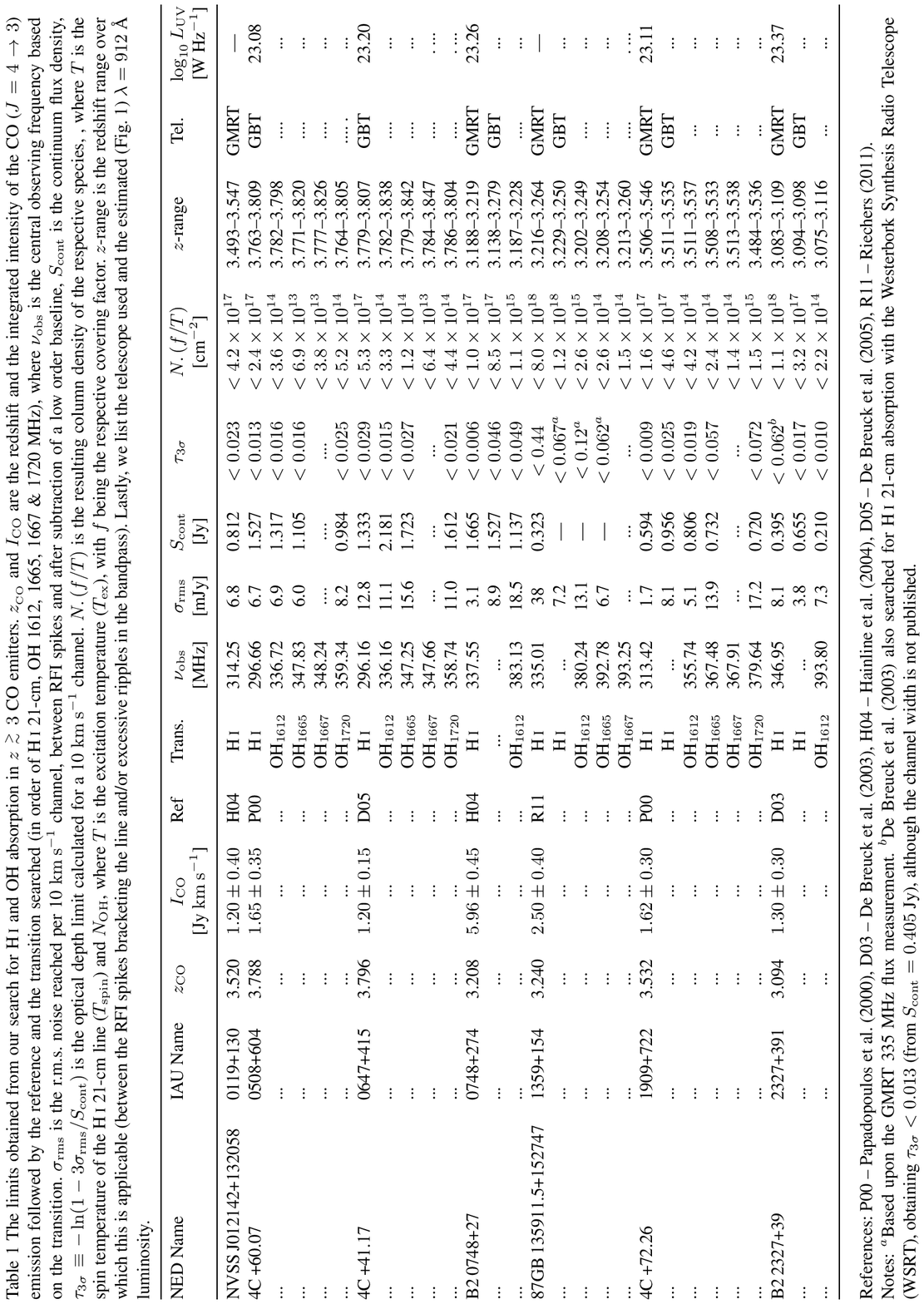}}
\end{picture}
%\caption{} % CHECK THIS DOESN'T SHOW UP ANYWHERE, BUT COULD BE PUSHED OFF BOTTOM OF PAGE
\label{obs}
\end{figure*}

In Fig. \ref{spectra} we show the final spectra 
and summarise these in Table~\ref{obs}, from which we see neither \HI\ nor OH were detected in any of the targets.
Since the OH absorption in the five known redshifted systems (Sect. \ref{intro}) was detected on the basis of a previous
\HI\ absorption detection \citep{cps92,cry93,cmr+98,cdn99,kb03}, 
and that  the absorption strength is only expected to be $\lapp10^{-4}$ times that of the 21-cm absorption \citep{cdbw07}, we will
treat \HI\ as a prerequisite for OH absorption and thus focus our discussion around the 21-cm results.
 
\begin{figure*}
\centering \includegraphics[angle=-90,scale=0.66]{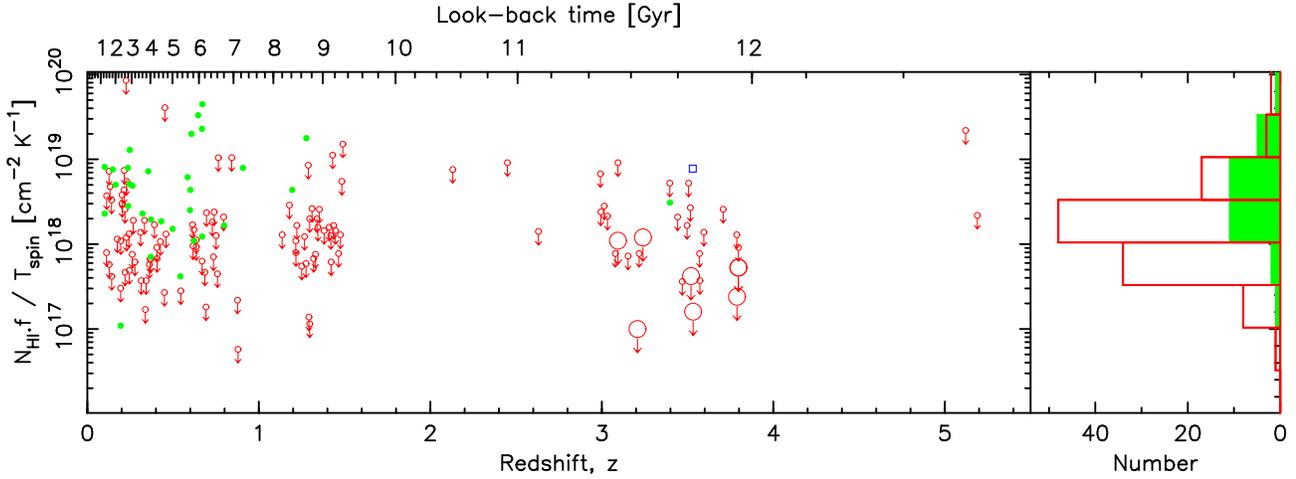}
\caption{The line strength ($1.823\times10^{18}\int\tau\, dv$) versus redshift for the $z\geq0.1$ \HI\ 21-cm
absorption searches. The filled circles/histogram represent the detections and the unfilled circles/histogram the $3\sigma$ upper limits to the non-detections,  where the large circles signify the CO emitters searched for 21-cm absorption. The unfilled square at $z=3.53$ shows the tentative detection of \citet{akk16}, which we will treat as a non-detection
  until confirmed.}
\label{N-z}
\end{figure*}

\begin{figure*}
\centering \includegraphics[angle=-90,scale=0.66]{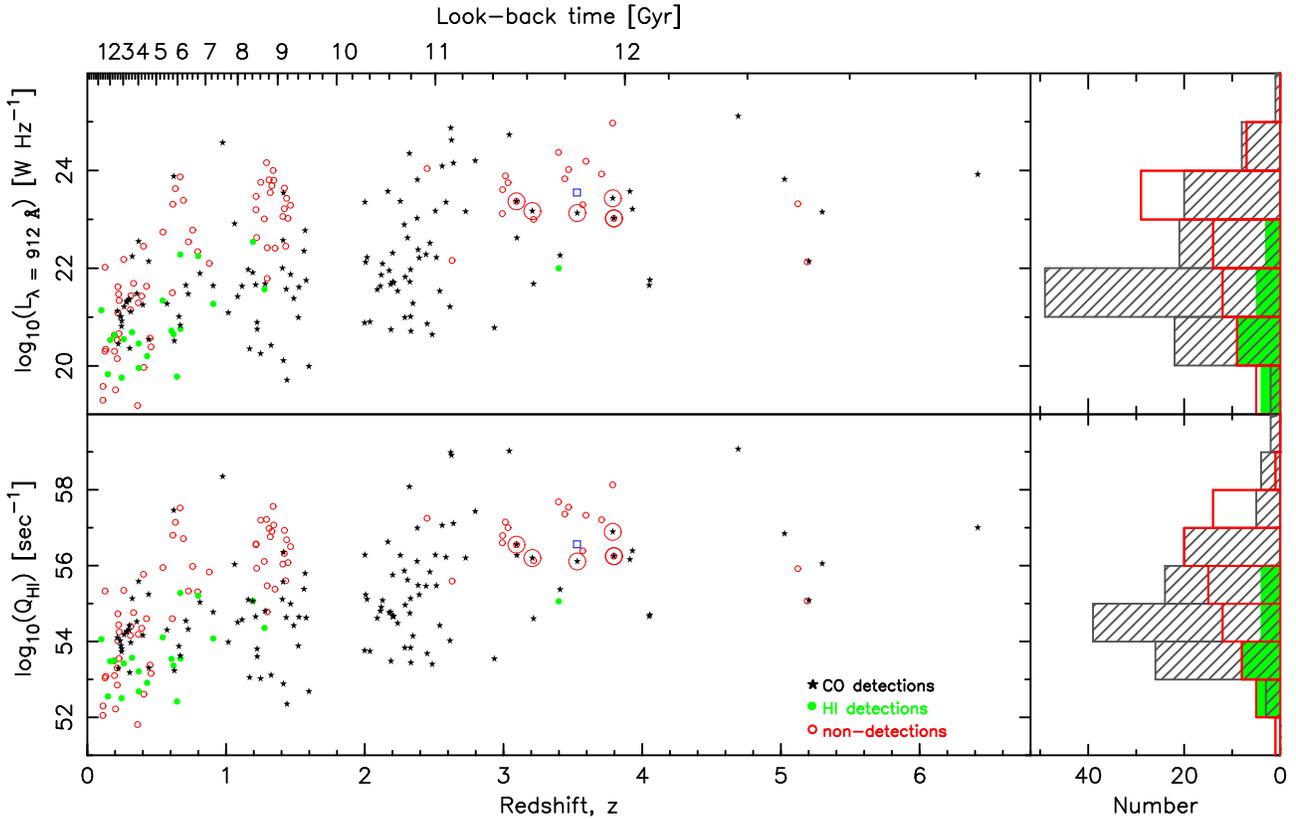}
\caption{The $\lambda = 912$ \AA\ continuum luminosity (top) and ionising ($\lambda \leq912$ \AA) photon rate (bottom)
  versus redshift for the sources detected in CO emission (filled stars) in addition to those searched in \HI\ 21-cm
  absorption (filled circles - detections, unfilled circles - non-detections, where the large circles signify the CO
  emitters searched). Again, the unfilled square shows the tentative detection of \citet{akk16}, where $L_{\rm UV} =
  3.5\times10^{23}$ \WpHz\ and $Q_\text{\HI} = 3.6 \times10^{56}$~sec$^{-1}$. The hatched histogram shows the distribution for the CO emitters, the filled the 21-cm detections and
  the unfilled the non-detections.}
\label{L-ion}
\end{figure*}

From Fig. \ref{N-z}, we see that our survey is as sensitive as many of the previous surveys for redshifted associated \HI\ 21-cm absorption
and deep enough to re-detect the majority of the known 21-cm absorbers.\footnote{\label{hi_ref}Compiled from \citet{mir89,vke+89,ubc91,cps92,cmr+98,mcm98,ptc99,ptf+00,mot+01,ida03,vpt+03,cwm+06,gs06,gss+06,ssm+10,css11,cwm+10,cwwa11,cwt+12,gmmo14,akk16} with the non-detections compiled in \citet{cw10}. }
Therefore, we may expect $\approx2$ detections from our sample, or $\approx7$ from all the previous searches at
$z\gapp3$, based upon the 27\% detection rate at $z\lapp3$.  However, there is just one detection 
at  $z\gapp3$ (\citealt{ubc91}), which is the only source below the critical UV luminosity (with 
$L_{\rm UV}\approx3\times10^{22}$ \WpHz). Due to all of our targets being at
$z\gapp3$, despite our best efforts, the UV luminosity of each of these was estimated to be close to the critical value
of $L_{\rm UV}\sim10^{23}$ \WpHz. This is seen in Fig.~\ref{L-ion}, which shows the distribution for the \HI\ 21-cm
searches and the CO detections.\footnote{UV luminosities and ionising photon rates could only be estimated for five of
  the targets, although the remaining two, 0119+130 \& 1359+154, could have luminosities below the critical value
  (Fig. \ref{SEDs_full}).}
From the figure it is clear that, while \HI\ 21-cm is never detected above $L_{\rm UV}\sim10^{23}$ \WpHz\
($Q_\text{\HI}\sim10^{56}$~sec$^{-1}$), CO emission is detected in several cases, up to luminosities as high as
$L_{\rm UV}= 1.3\times10^{25}$ \WpHz\ ($Q_\text{\HI}\ = 1.2\times10^{59}$~sec$^{-1}$), which is over two orders of
magnitude more luminous than the critical \HI\ value.\footnote{Unfortunately, the CO non-detections are generally not
  published, although of the eight non-detections of \citet{enf+14}, rest-frame UV photometry exists for three, one of
  which has a  luminosity above the critical value, with two below. Hence, like the incidence of \HI\ 21-cm absorption,
  there must be another effect at play (e.g. gas disk orientation) in the detection of CO emission.}  
We now discuss possible reasons why CO emission
is readily detected at ionising photon rates which are inhospitable to the neutral atomic gas.

\subsection{Possible reasons for the presence of CO above the critical photo-ionising rate}

\subsubsection{Shielding by dust}
\label{dust}

Interstellar molecules form on dust grains and so shielding by dust may be a reason why CO is detected at
ultra-violet luminosities which ionise the \HI. However, the fact that the UV/visible fluxes are high may suggest
that the sight-lines towards the sources have relatively low dust obscuration. 
 \begin{figure}
\centering \includegraphics[angle=-90,scale=0.44]{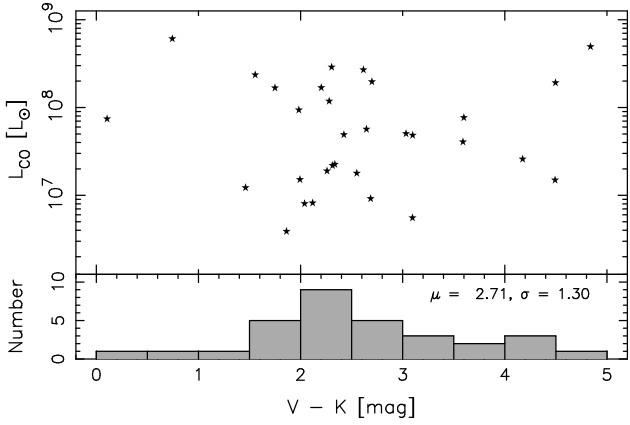}
\caption{The luminosity of the CO emission versus the optical--near infrared colours of the CO emitters, for which both
  $V$ and $K$ magnitudes are available.}
\label{V-K}
\end{figure}
In order to address this, in Fig.~\ref{V-K} we show the optical--near infrared colours of the CO emitters as a proxy for dust obscuration. 
From these, we see that the colours span a very similar range to optically selected
QSOs (\citealt{shr+07}, which have a mean and a standard deviation of $\mu=2.92$ and $\sigma = 0.75$, respectively) in comparison to $V-K\approx 5 - 9$ for the five known redshifted molecular absorption systems (see figure~1 of \citealt{cwc+11}) and $V-K\gapp2$ for associated
21-cm absorption \citep{cw10}.  

The molecular absorption is due to low-excitation gas ($T_{\rm CMB}\lapp T_{\rm ex}\lapp10$ K,
\citealt{wc95,wc96,wc96b,wc97}), where the density is insufficient to populate the higher levels of the molecules,
according to the higher kinetic temperatures (i.e. $T_{\rm kin}>> T_{\rm ex}$ ). For gas in emission, however, both the
volumic density excitation temperature must be higher, although lower than the kinetic temperature ($T_{\rm kin}
\approx50 - 100$~K, cf. $T_{\rm ex}\approx13$~K, \citealt{prv+00,piv07}), indicating that the gas is, on average, at
intermediate densities and perhaps less obscured by dust, as is indicated by the less reddened colours.

However, the colours may not be a reliable indicator of the actual column densities, if the dense regions are embedded
in a diffuse medium. Very dense self-shielded clumps of high column density gas could give rise to substantial CO
emission with un-reddened colours, if the surface filling factor of the dense gas is low.  These low filling factors
($\lapp1$\% in absorption, \citealt{pc94,pcm94}) offer an alternate explanation to the lack of 21-cm
absorption. However, as seen from Fig. \ref{N-z}, 21-cm absorption is readily detected at $z\lapp1$ and is also detected
in all five of the decimetre/millimetre band molecular absorbers \citep{cps92,cry93,cmr+98,cdn99,kb03} and so low
filling factors are unlikely to be the cause of the 21-cm non-detections.
 
\subsubsection{Self-shielding}

Another viable reason for the  detection of CO at UV luminosities unfavourable to 21-cm absorption
 is the possibility that the molecular gas is self-shielded, due to its higher density. 
\citet{cw12} showed that an ionising photon rate of $Q_\text{\HI}\sim10^{56}$~sec$^{-1}$ was sufficient
to ionise all of the neutral gas in a galaxy with the \HI\ distribution of the Milky Way \citep{kk09}.
 This critical value is therefore valid for a central gas density of $n_0\approx 10$~cm$^{-3}$, which decays with galactocentric
radius with a scale-length of $R=3$~kpc. However, CO traces gas of significantly higher density, and applying the canonical
$n_0\sim10^3$~cm$^{-3}$ to the maximum luminosity at which CO is detected, $L_{\rm UV}= 1.3\times10^{25}$ \WpHz,
gives a scale-length of $R\approx500$ pc and a total gas mass of $M_{\rm gas} =\int_0^{r}\rho\, dV =  4\times10^{9}$\,\Mo, both values being typical of a spiral galaxy (e.g. \citealt{ckb08}).
%X-n-Mcyl_r_overlay

However, while an exponential disk may reasonably model the large-scale neutral atomic gas distribution in a galaxy,
the denser molecular gas is expected to be embedded in discrete clouds within this.  This could mean that the  
molecular gas is shielded by the enveloping atomic gas (e.g. \citealt{kmt08,kmt09,mk10}), since the dissociation of \MOLH\ requires higher
energy photons than the ionisation of \HI\ (14.7 cf. 13.6 eV, \citealt{spi48}).\footnote{Since \MOLH\ is homo-nuclear it has no dipole moment, making the dissociation of \MOLH\ by 4.52 eV photons forbidden.}
Because of its large dipole moment, it is CO, not \MOLH\ directly, which is detected in these millimetre band surveys and so it is possible 
that CO emission at high redshift could arise in very different environments than in the Galaxy, and
use of the Galactic conversion ratio \citep{sbd+88} 
would not be warranted. Therefore, since we have no direct
measure of the \MOLH\ emission properties of these objects, it is more prudent to consider the CO directly. 
This is dissociated at 11.09 eV (e.g. \citealt{vvb09}), which would suggest that the CO actually shields the \HI, and 
being readily detected, while \HI\ is not, this would indicate that the CO may be more remote from the
AGN activity, a possibility which we now discuss.

\subsubsection{CO remote from the UV continuum}

CO emission in high redshift galaxies has been observed to extend to large scales; $\gapp10$ kpc in the $4\rightarrow3$ transition \citep{prv+00,isa+12} and $\gapp50$ kpc
in the $1\rightarrow0$ transition. This extended emission appears to arise from
external components such as merging gas-rich galaxies \citep{dno03,dnm+03,ddn+05,efr+13} and
Ly\AL\ halos around the host galaxy \citep{nnd+09}. Furthermore, the CO emission is often aligned with the  radio
jet axis \citep{kesh04,nnd+09}, which is a further indication that the CO is not confined to the host radio galaxy.

\begin{figure}
\centering \includegraphics[angle=-90,scale=0.48]{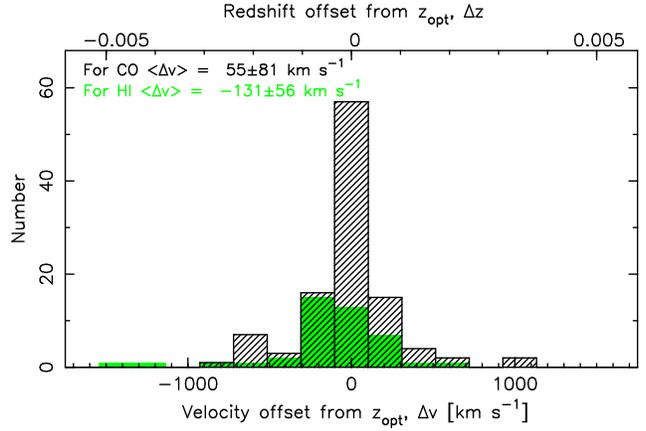}
\caption{The velocity offset from the systemic for the CO emission (hatched histogram) and \HI\ absorption (coloured histogram) detections.}
\label{dv-hist}
\end{figure}

Since the \HI\ absorption is believed to arise in the disk, if not the obscuring nuclear material \citep{cw10}, the
molecular gas detected in emission may therefore be spatially offset from the atomic detected in absorption.  If this
were the case, we may expect differences in the offsets of each species from the systemic redshift.  Calculating these
via
 \[
\Delta v = c \frac{(\Delta z + 1)^2 -1}{(\Delta z + 1)^2 +1}\approx c\,\Delta z,  {\rm ~where~}\Delta z = \frac{z_{_{\rm opt}}+1}{z_{_\text{CO,HI}}+1} - 1,
\]
gives mean offsets of $\left<\Delta v \right> =  55\pm81$ \kms\ for the CO emission and $\left<\Delta v \right> = -131\pm56$ \kms\
for the \HI\ absorption (Fig.~\ref{dv-hist}). 
That is, on average, the \HI\ absorption may be more offset from the systemic velocity, where we would expect this
to be less, if the \HI\ were intrinsic.  
%awk < HI_FWHM.dat '{if ($1 !~/#/) print $0}' | wc -l   43
%awk < HI_FWHM.dat '{if ($1 !~/#/ && $8 == "vpt+03") print $0}' | wc -l
A Kolmogorov-Smirnov test of the $\Delta v$ values gives a probability of $9.81\times10^{-3}$ that the two samples are
drawn from the same population, which is significant at $2.58\sigma$ assuming Gaussian statistics (Fig. \ref{CO-UV}). 

This overall blueshift of the cool neutral gas was also noted by \citet{vpt+03}, whose sample comprises 19 of the 43
\HI\ absorbers for which we can estimate $\Delta v$.  The blueshift was attributed to outflowing gas
(e.g. \citealt{msc+15} and references therein) and fast outflows could account for the offsets of $\Delta v\gapp500$
\kms, which exceed the escape velocity of a large spiral galaxy. Fast outflowing gas is predominant in compact radio
sources (e.g. \citealt{vpt+03,gmmo14}), which may have higher \HI\ detection rates than in the extended sources
(e.g. \citealt{pcv03,gs06a}). Since such kinetic feedback can deplete the hydrogen in galaxies
(e.g. \citealt{dc12,cms+14,acm+16}), the depletion of gas in more mature (extended) objects could explain the paucity
of 21-cm absorption detections, although high resolution radio imaging of the CO emitters would be required in order to
verify this. Note, however, that once the $L_{\rm UV}\sim10^{23}$ \WpHz\ sources are removed from the radio source
samples, the \HI\ detection rates in compact objects does not differ significantly from those in non-compact objects,
which is consistent with the compact objects being less mature and luminous than the non-compact sources
\citep{cw10,ace+12}.  Thus, we believe that photo-ionisation of the neutral gas remains a key factor in the  non-detection of \HI.

% awk < HI_FWHM.dat '{if ($1 !~/#/) print $0}' | sort -n -k 4   JUST FOUR OUT OF 43 WITH Z > 1

\begin{figure}
\centering \includegraphics[angle=-90,scale=0.50]{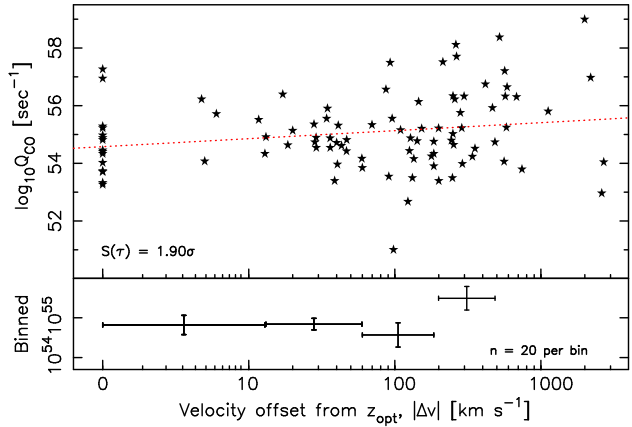}
\caption{The CO photo-dissociation rate versus the absolute velocity offset from the systemic for the CO emission. $Q_\text{CO}$ is obtained from the  SED fits at $\lambda \leq1119$ \AA, cf. $\lambda \leq912$ \AA\ for $Q_\text{\HI}$, where possible.}
\label{Q_DIS}
\end{figure}

Furthermore, in drawing any conclusions from the velocity offsets, it should be noted that many of the optical redshifts
are only accurate to the second decimal place giving uncertainties of $\sim10^2$ \kms\ and so the absolute value of
$\Delta v$ may be unreliable, with many of the $\Delta v=0$ values arising by default (see Fig.~\ref{Q_DIS}). The
narrower spread in \HI\ offsets\footnote{An F-test rejects the null hypothesis of equal variances in the \HI\ and
  CO samples at a very high significance (a $p$-value of $6.65\times10^{-9}$).} may suggest that the atomic gas is more likely
to be associated with the radio source, although one can not draw any firm conclusions from the distribution of the
offsets until more accurate optical redshifts become available.

Returning to the dissociation of CO by the UV flux, 
in Fig.~\ref{Q_DIS} we show the CO photo-dissociation rate  versus $\Delta v$, from which we only see a weak correlation.
Assuming that this is diluted by inaccurate optical redshifts and projection effects, and provided that
$\Delta v$ is a reliable proxy for the distance between the AGN host, 
this may be evidence of the molecular clouds being able to survive larger rates
($Q_\text{CO}\gg10^{56}$~sec$^{-1}$) due to their larger distances from the AGN.

\subsubsection{UV continuum arising from star formation}
\label{sf}
Another possibility for the detection of CO emission in the brightest UV sources, is that both luminosities are correlated with
star forming activity: Due to the low efficiency of star-formation, the consumption of molecular gas 
is far from complete, with a depletion times of $2-3$~Gyr at $z=0$ \citep{blw+08} compared  to 0.7 Gyr at $z=1-3$ \citep{tng+13}. That is,
there is an apparent increase in star formation efficiency with redshift.

\begin{figure}
\centering \includegraphics[angle=-90,scale=0.51]{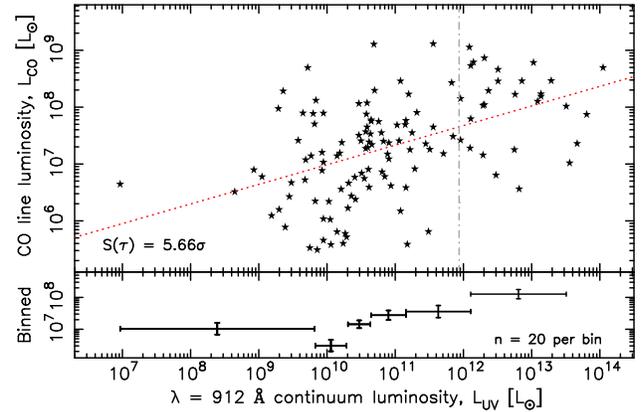} 
\caption{The CO line versus the UV continuum luminosity for the CO emitters for which this can be determined (see \citealt{cwsb12}). The vertical broken line at $L_{\rm UV} = 8.6\times10^{11}$ \Lo\  shows $L_{\rm UV} = 10^{23}$ \WpHz\ and the dotted line the fit to all of the points.}
\label{CO-UV}
\end{figure}

Investigating the origin of the UV luminosity, in  Fig. \ref{CO-UV} we show the CO line luminosity versus the UV
continuum luminosity, from which a non-parametric generalised Kendall-tau test gives a $P(\tau) =
1.5\times10^{-8}$ probability of the observed correlation occuring by chance, which is significant at $5.66\sigma$.
Such a relationship may be expected,  since far-infrared emission is a tracer of the star formation rate (SFR) in Ultra
Luminous Infrared Galaxies (ULIRGS, where $L_{\rm FIR}\gapp10^{12}$\,\Lo, e.g. \citealt{ken98}) and six of our sample are
estimated to have $L_{\rm FIR}\approx5 - 14\times10^{12}$\,\Lo\ (Fig.~\ref{SEDs_full}). The fact that they have both
high CO and FIR luminosities, suggests that they are host to high star formation rates
(e.g. \citealt{fct+13,gtl+15}). However, it is generally assumed that all of the FIR emission arises from star formation
and such a large dynamic range may be subject to selection effects, such as an increasing luminosity with redshift probing
differing host morphologies and amplification of the flux by gravitational lensing.\footnote{Of our sample, 0748+27 and
1359+154 are known to be lensed.}

\begin{figure}
\centering \includegraphics[angle=-90,scale=0.51]{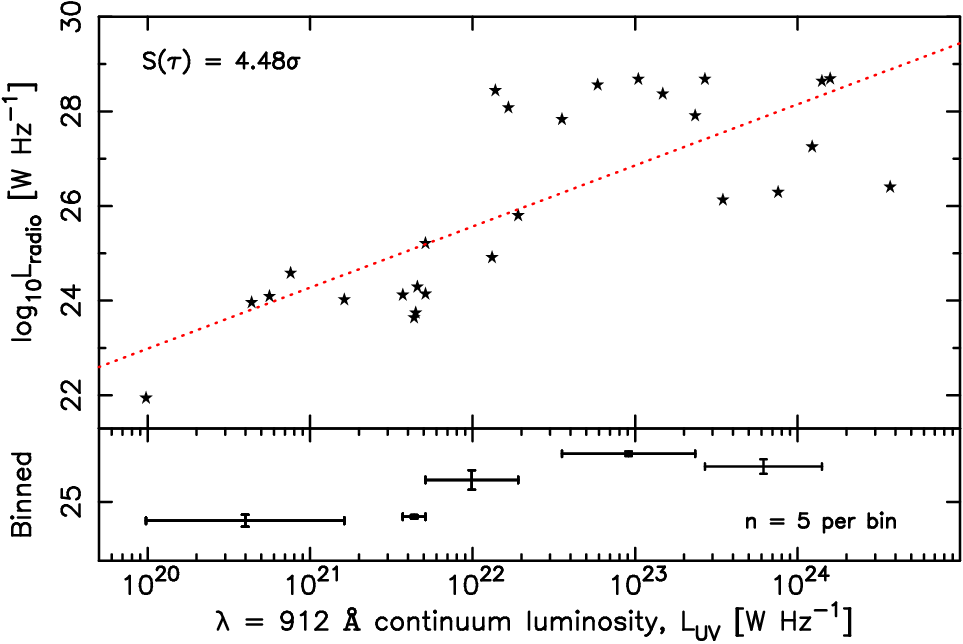}
\caption{The 1.4 GHz radio continuum versus that of the UV for the CO emitters, where available. 
For the $z\gapp0.1$ associated 21-cm absorbers, the radio powers are typically $L_{\rm radio}\gapp 10^{26}$ \WpHz\ (\citealt{cww+08}, or $\int L_{\rm radio}\,d\nu \gapp10^{35}$~W, \citealt{cwt+12}).}
\label{radio-UV}
\end{figure}

In Fig. \ref{radio-UV}, we also find a strong correlation between the radio and UV continuum luminosities ($P(\tau) =
7.32\times10^{-6}$ for the 27 sources for which the radio luminosity can be obtained). 
Since the radio--FIR correlation is often attributed to star-formation, this could strengthen the argument that the high UV
luminosities arise from star formation, although the radio--FIR relationship is generally limited to $L_{\rm radio}\lapp
10^{24}$ \WpHz\ (e.g. \citealt{sm96,bwb+15,ped+15}).
Furthermore, at $L_{\rm radio}\gapp 10^{23}$ \WpHz, the population rapidly becomes dominated by AGN (e.g. \citealt{ms07}), although this is only measured to $z\lapp0.3$ and it is not known how this
extrapolates to high redshift.
\begin{figure}
\centering \includegraphics[angle=-90,scale=0.49]{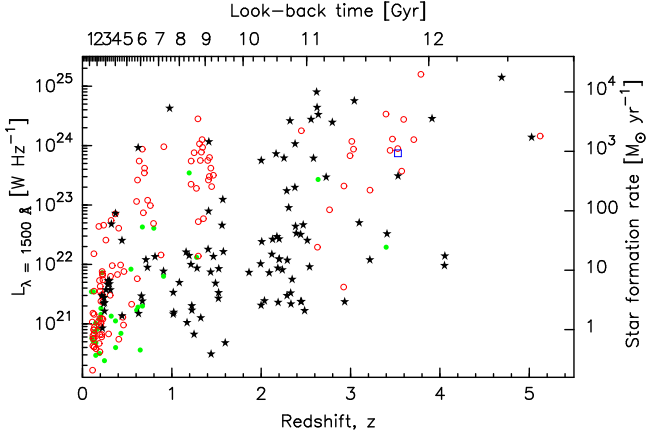}
\caption{The rest-frame $\lambda =1500$ \AA\ luminosity of the CO emitters (stars) and 21-cm absorption searches (circles), where the right axis shows the star formation rate
derived from $\text{SFR} = L_{\lambda =1500 \text{ \AA }}/8.0\times10^{20}$ \citep{mpd98}.}
\label{mpd}
\end{figure}

In Fig. \ref{mpd} we show the estimated star formation rates, based upon the UV ($\lambda =1500$ \AA) luminosities
\citep{mpd98}.  The maximum $L_{\lambda =1500 \text{ \AA }} = 3.5\times10^{23}$ \WpHz\ at which 21-cm is detected
corresponds to a SFR of $430$\,\Moy, with the maximum $L_{\lambda =1500 \text{ \AA }} = 1.4\times10^{25}$ \WpHz\ at
which CO is detected corresponding to 17\,000 \,\Moy.  By assuming that all of the FIR emission is due to star
formation, rates of $3000$ \Moy\ have been found in high redshift ($z\gapp6$) CO emitters (e.g. \citealt{bcn+03}),
although these rates are believed to occur in short bursts ($<10$ Myr).  Thus, we obtain extremely high star formation
rates based upon the UV luminosities of the CO emitters and hence believe that, at least partially, the high
luminosities, and subsequent dust heating, are caused by increased AGN activity (e.g. \citealt{gc00}).
 
One way to address the origin of the UV radiation would be through spectral classification, although, of the six of the seven target
spectra which could be found, all are UV (rest) emission line spectra, with little continuum, and so are difficult to classify. However,
NED classifies three as QSOs (0748+27, 1359+154 \& 2327+39) and four as galaxies (0119+130, 0508+604, 0647+415 \& 1909+722),
although the 50\% detection rate in both type-1 and type-2 objects \citep{cw10} suggests that the host morphology
makes little difference to the detection of 21-cm absorption in objects below the critical UV luminosity.
 
\section{Summary}

We have undertaken a survey for atomic (\HI\ 21-cm) and molecular (OH 18-cm) absorption in seven strong CO
emitters at $z\gapp3$. Our motivation was to circumvent the need for an optical redshift, to which to tune the receiver, which
selects against the detection of decimetre absorption at high redshift. Tuning to the CO redshift also has the advantage
that large amounts ($\gapp10^{10}$\,\Mo) of molecular gas (e.g. \citealt{dnm+03,sv05}) are known to
exist and \HI\ absorption and CO emission are often coincident in low-z active galactic nuclei (e.g. \citealt{crjb98},
cf. \citealt{kor96} and \citealt{cjhb99}, cf. \citealt{afm+87}).  Despite this, no absorption of either species was detected.
We attribute this as being due to the targets, despite careful selection, having rest-frame ultra-violet luminosities close 
to the value above which all of the neutral gas is ionised \citep{cw12}. 

While the non-detection of \HI\ 21-cm absorption within the host galaxies of sources with ionising photon rates
exceeding $Q_\text{\HI}\approx3\times10^{56}$~sec$^{-1}$ ($L_{\rm UV} \gapp 10^{23}$ \WpHz) appears to be a universal
phenomenon, holding true for various heterogeneous and unbiased samples, with varying selection criteria and over all redshifts, we
find that CO emission is readily detected at $L_{\rm UV} \gg10^{23}$ \WpHz.  In order to explain the detection of
molecular gas in emission, while atomic gas in absorption remains undetected, we suggest the following, non-mutually
exclusive, possibilities:
\begin{enumerate}
\item That, due to the ionisation of \HI\ occuring at a lower energy than the dissociation of \MOLH, as per Galactic clouds, the molecular gas is shielded by the atomic gas.
However, by the same argument the CO should shield the \HI, although this is likely to be embedded in dense molecular cores which are enveloped
by the more diffuse atomic gas.
\item CO traces higher density gas ($n_{\text{H}_2}\gapp10^3$~\ccm, cf. $n_{\text{\HI}}\sim10$~cm$^{-3}$ for \HI), which
  could provide the CO with better self-shielding against the UV flux. From our model galaxy \citep{cw12}, we find that
  the maximum luminosity at which CO is detected ($L_{\rm UV}= 1.3\times10^{25}$ \WpHz) gives a gas
  density distribution and  mass which is consistent with the molecular gas  in a typical spiral galaxy. This, however, relies on the
  assumption that the distribution of CO is spatially coincident with the \HI, when the molecular gas  is likely to be embedded in discrete
  clumps.
\item The CO gas is located in a dustier environment which provides better shielding against photo-ionisation. This
  appears to be the case for both molecular and \HI\ 21-cm {\em absorption} which are correlated with the optical--near
  infrared colours of the sight-line. For the CO {\em emission}, however, no correlation is seen and this is readily
  detected at colours which are similar to those of optically selected QSOs, ($V-K\lapp5$), whereas CO {\em absorption}
  has thus far only been detected towards much redder sight-lines ($V-K\gapp5$). This suggests either that dust does not play as an important role 
  in CO emission as in absorption or that the CO is physically offset from the sight-line (to the radio/UV source) along
which the colours are measured.
 \item  At high redshift, there is evidence that the CO emission is physically remote from the source of the continuum emission, 
meaning that the CO is subject to much lower UV fluxes than the cool, neutral gas located within the host galaxy.
\end{enumerate}
We also investigate the possibility that the UV emission arises primarily from star formation, as may be evident through
a strong correlation between the UV and CO luminosities. However, it is not clear how much of the ultra-violet emission
is due to nuclear activity in these objects, although three of our seven targets  appear to be QSOs and all seven have radio luminosities typical
of AGN. Thus, it is possible that a large fraction of UV emission does not arise from star formation and that
the presence of CO is these UV luminous sources could be explained by self-shielding of the molecular gas or their
distance from the continuum source.

Whatever the cause, the detection of \HI\ and OH absorption at high redshift remains elusive, with wide-band decimetre-wave spectral 
scans of optically obscured objects being required to detect these tracers of cool gas at high redshift and perhaps shed light upon the paucity of the
star forming reservoir in the locale of warm molecular gas.

\section*{Acknowledgements}
We wish to thank Bjorn Emonts for his valuable input as well as the staff of the GBT and GMRT that made these observations possible. The GMRT is run by the National Centre for Radio
Astrophysics of the Tata Institute of Fundamental Research.
The National Radio Astronomy Observatory is a facility of the National Science Foundation operated under cooperative agreement by Associated Universities, Inc.
This research has made use of the NASA/IPAC Extragalactic Database (NED) which is operated by the Jet Propulsion
 Laboratory, California Institute of Technology, under contract with the National Aeronautics and Space Administration and 
NASA's Astrophysics Data System Bibliographic Service.
This research was conducted by the Australian Research Council Centre of Excellence for All-sky Astrophysics (CAASTRO), through project number CE110001020.

%\bibliographystyle{../mn2e}  
%\bibliography{aa,ref}

\label{lastpage}

\end{document}